Capillary condensation in cylindrical nanopores

Silvina M. Gatica[†] and Milton W. Cole
Department of Physics and Materials Research Institute
104 Davey Laboratory
Pennsylvania State University
University Park, PA 16802

[†]Corresponding author. Current address:
School of Computational Science, MS 5C3
George Mason University
Fairfax, VA  22030
sgatica@gmu.edu

Abstract

Using grand canonical Monte Carlo simulations, we have explored the phenomenon of capillary condensation (CC) of Ar at the triple temperature inside infinitely long, cylindrical pores. Pores of radius R= 1 nm, 1.7 nm and 2.5 nm have been investigated, using a gas-surface interaction potential parameterized by the well-depth D of the gas on a planar surface made of the same material as that comprising the porous host. For strongly attractive situations, i.e., large D, one or more (depending on R) Ar layers adsorb successively before liquid fills the pore. For very small values of D, in contrast, negligible adsorption occurs at any pressure P below saturated vapor pressure $P_0$; above saturation, there eventually occurs a threshold value of P at which the coverage jumps from empty to full, nearly discontinuously. Hysteresis is found to occur in the simulation data whenever abrupt CC occurs, i.e. for R≥ 1.7 nm, and for small D when R=1nm. Then, the pore-emptying branch of the adsorption isotherm exhibits larger N than the pore-filling branch, as is known from many experiments and simulation studies. The relation between CC and wetting on planar surfaces is discussed in terms of a threshold value of D, which is about one-half of the value found for the wetting threshold on a planar surface. This finding is consistent with a simple thermodynamic model of the wetting transition developed previously.

1. Introduction:

The subject of capillary condensation has stimulated research concerning its origin for at least three centuries [1-3]. Newton's monograph "Opticks" described experiments of F. Hauksbee which revealed that the capillary rise of a liquid in a tube varies inversely with the tube radius, R. Laplace recognized that this phenomenon is a consequence of the pressure discontinuity at the liquid-vapor interface. Using contemporary ideas of Young (concerning wetting behavior), Laplace developed a relationship between the capillary rise and the three relevant surface tensions. Since these same surface tensions determine the contact angle of the liquid drop on a flat surface, Laplace was able to establish a direct relationship between the wetting behavior and the phenomenon of capillary condensation. A corollary to this finding (in the case of a complete wetting film) is the

Kelvin equation, which expresses the value of the chemical potential μ at which CC occurs in terms of the radius of the tube.

A significant body of modern research is devoted to the study of fluid imbibition within nanopores [1-9]. Since the Laplace/Kelvin analysis is based on *macroscopic* properties of the fluid, omitting the wall-fluid interaction, it is not quantitatively trustworthy in this small pore regime. Thus, numerous investigations have been carried out to explore the CC behavior in such nanopores [3-9]. Virtually every analytical and numerical method used for fluids (computer simulation, density functional methods, lattice gases, simple thermodynamic models…) has been employed in these investigations. The present study is focused on a specific kind of problem: we look for trends in the CC behavior as a function of the gas-surface interaction potential. This investigation parallels our recent work concerning wetting transitions on planar surfaces [10,11]. In these studies, the grand canonical Monte Carlo method is used to determine how the adsorption behavior varies with the external potential. As expected, when the gas-surface interaction is very attractive, CC occurs readily at a low value of μ. In contrast, when the interaction is weakly attractive, little adsorption occurs for either planar or cylindrical geometries. By performing a set of such calculations, we establish the relationship between the threshold attraction required for CC to occur and that required for complete wetting on a flat surface. Similar, but more limited, studies of the dependence of CC on interaction strength have been carried out by Gubbins's group [3,5].

The outline of this paper is the following. Section 2 describes the geometry, interactions and computational methods used in this research. Section 3 describes our results (densities and isotherms) and compares them to the behavior found on planar surfaces. Section 4 summarizes our conclusions. We note in advance that the parameter space for this problem is enormous, including a range of radius (R) values, thermodynamic conditions and interaction strengths; our goal is to explore a limited subset of these problems in sufficient depth to elucidate the key problem: what is the evolution of CC behavior as the strength of the gas-surface interaction is varied? Here, we have investigated Ar at its triple temperature, for specificity. Because the results found for the CC behavior are semiquantitatively consistent with those derived from a so-called "simple model" of adsorption, we believe that the model can be used to extrapolate our findings to other simple fluids near their respective triple points.

2. Interactions and computational method

In computer simulation studies of adsorption, one needs to specify the gas-gas and gas-surface interactions. For the Ar-Ar interaction, we assume that the pair potential U at separation $r_{12}$ is of the Lennard-Jones (LJ) 6-12 form:

$$U(r_{12}) = 4\,\varepsilon_{Ar}\,[(\sigma_{Ar}/r_{12})^{12} - (\sigma_{Ar}/r_{12})^{6}] \qquad [1]$$

Here, the potential parameters are given the venerable values $\varepsilon_{Ar}$ =120 K and $\sigma_{Ar}$ =3.41 Å, respectively. We assume for simplicity that the host material surrounding the pore is a uniform continuum, with constant atomic number density $n_p$. We postulate that the gas-surface interaction V(**r**) is a sum of Ar-substrate atom pair interactions, each of LJ form, with (adjustable) parameters $\varepsilon$ and $\sigma$. The functional form of V(**r**) depends on the geometry; in the present case we will be considering planar and cylindrical cases. Upon integration over a half-space continuum, in the case of the planar geometry, one obtains a net adsorption potential $V_{planar}(z)$ that can be expressed in terms of the planar surface well-depth D:

$$V_{planar}(z)/D = (4/27) (d/z)^9 - (d/z)^3 \quad [2]$$

Here, z is the normal distance of the Ar atom from the substrate and the length $d=(C/D)^{1/3}$ depends on the system. We note that $C= (2\pi/3) n_p \varepsilon \sigma^6$ is the asymptotic gas-surface van der Waals interaction coefficient, defined by the limiting behavior $V \sim -C/z^3$. The potential in Eq. 2 has a minimum at $z_{min} = (2/3)^{1/3}$ d. The 3-9 potential of Eq. 2 provides a qualitative description of gas-surface interactions that is analogous to the role played by the 6-12 LJ gas-gas interaction in calculations involving the statistical mechanics of fluids. Results found for such simple forms should be at least qualitatively consistent with results derived from more accurate potentials.

Now, let us turn to the case of a cylindrical pore of radius R. We write the interaction as a sum of attractive and repulsive terms:

$$V(r) = V_{repul}(r) + V_{attr}(r)$$

Here, we compute the attractive interaction by integrating over the surrounding medium (|**r**'|>R) the same pair attraction used in the planar case:

$$V_{attr}(r) = - 4 n_p \varepsilon \sigma^6 \int d\mathbf{r}' \, |\mathbf{r} - \mathbf{r}'|^{-6}$$

$$V_{attr}(r) = - (6 C/\pi) \int d\mathbf{r}' \, |\mathbf{r} - \mathbf{r}'|^{-6}$$

The result of this integration is [11]

$$V_{attr}(r) = - (3 \pi/2) (C/R^3) F(3/2, 5/2; 1; y^2) \quad [3]$$

Here F is a hypergeometric function and y=r/R is the reduced distance of the Ar atom from the axis of the pore. We employ a specific value of the coefficient C= 600 meV-Å$^3$, which is characteristic of the Ar interaction with alkali metals and insulators; the value of C for Ar interactions with graphite and metals is typically a factor of two or three higher [13]. With this value of C, the threshold value of D for Ar wetting at the triple point on a planar surface is $D_{wet}$ =53 meV [11]. This means that if D<53 meV, a wetting transition is expected to occur above the triple temperature.

For the repulsive part of the interaction, which is short-ranged, we assume that this part of the interaction coincides with that of the planar surface at the same separation z=R-r. We expect this approximation to not significantly affect our conclusions. From Eq. 2, therefore,

$$V_{repul}(r) = (4/27)(C^3/D^2)(R-r)^{-9} \quad [4]$$

Thus, our net potential is expressed in terms of parameters (C,D) used to characterize adsorption on planar surfaces.

Figure 1 shows the potential energy of the Ar atom for the case R=1.7 nm, with various assumed planar well-depths. Note that the well-depth in the cylindrical pore case exceeds D by about 10%, a result of the concave geometry. We note that D is measured, or calculated, to be 96, 72, 53, 36 and 21 meV for Ar on graphite, MgO, NaF, Mg and Li surfaces, respectively [13,14]. A 10% uncertainty is typical of these semiempirical values.

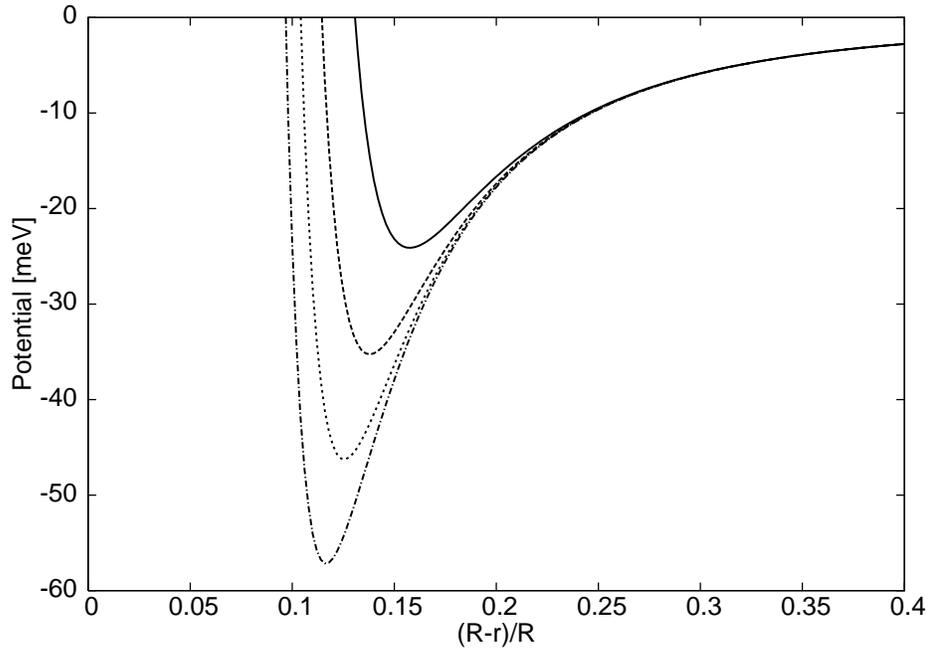

Fig. 1 The potential energy V(r) of an Ar atom in a pore of radius 1.7 nm, as a function of the distance from the pore's wall. The curves correspond to planar surface well depths D=50 meV (dashed-dotted), 40 meV (dotted), 30 meV (dashed) and 20 meV (solid).

The simulation method used here is the grand canonical Monte Carlo method. Apart from the cylindrical geometry, our computational procedure coincides with that used in our previous studies of wetting transitions on planar surfaces and on the external surface of a bundle of carbon nanotubes [11,15,16]. We have explored the effects of varying the length L of the unit cell (parallel to the pore axis) that is reproduced periodically in the computations. The choice L=8 nm was made because a negligible change in behavior was found when a larger value of L was used.

3. Simulation results

Figures 2 to 4 show adsorption isotherms in pores of radius 1 nm, 1.7 nm and 2.5 nm, respectively. In each case, the plotted quantity is a dimensionless density $N^* = N \sigma^3 / V_{cell}$, computed with the assumption that the full cross-sectional area ($\pi R^2$) is available for adsorption, i.e. $V_{cell} = \pi L R^2$. This represents an overestimate of the available volume because of the effect of the hard-core wall potential, which precludes significant adsorption in the outermost few Å near the wall. This reduced density region is apparent in Figs. 5 to 7, which show negligible (dimensionless) Ar density n(r) close to the pore wall; this "hard wall" exclusion of the density is essentially independent of P.

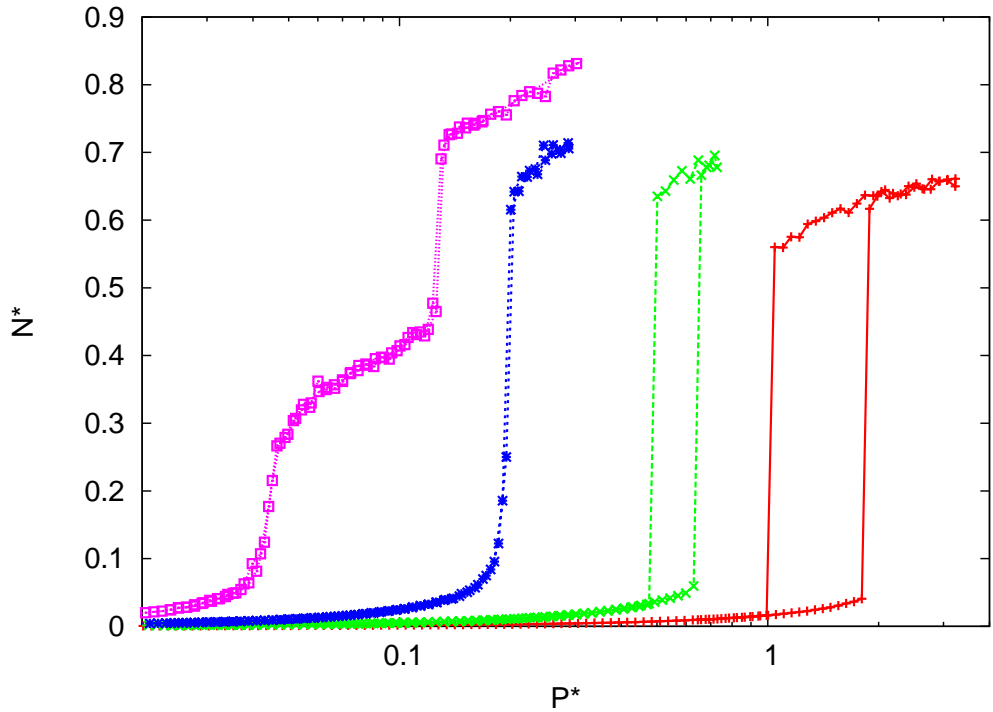

Fig. 2 (Color online) Reduced film density N* of adsorbed Ar atoms in pores of radius R=1 nm as a function of reduced vapor pressure $P^* = P/P_0$, where $P_0$ is the pressure at the triple point. Results have been computed with D=50 (squares), 40 (*), 30 (x) and 20 (+) meV. In each case, the adsorption branch is the one with higher N* at a given P. Curves for the two largest D values manifest no hysteresis.

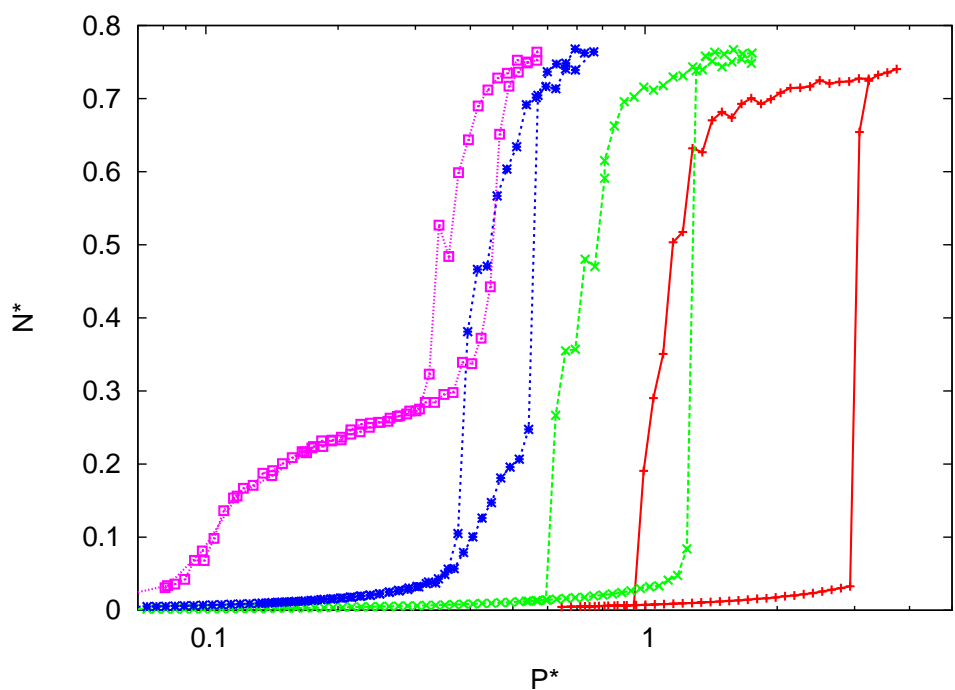

Fig. 3 (Color online) Same as Fig. 2 except R=1.7 nm

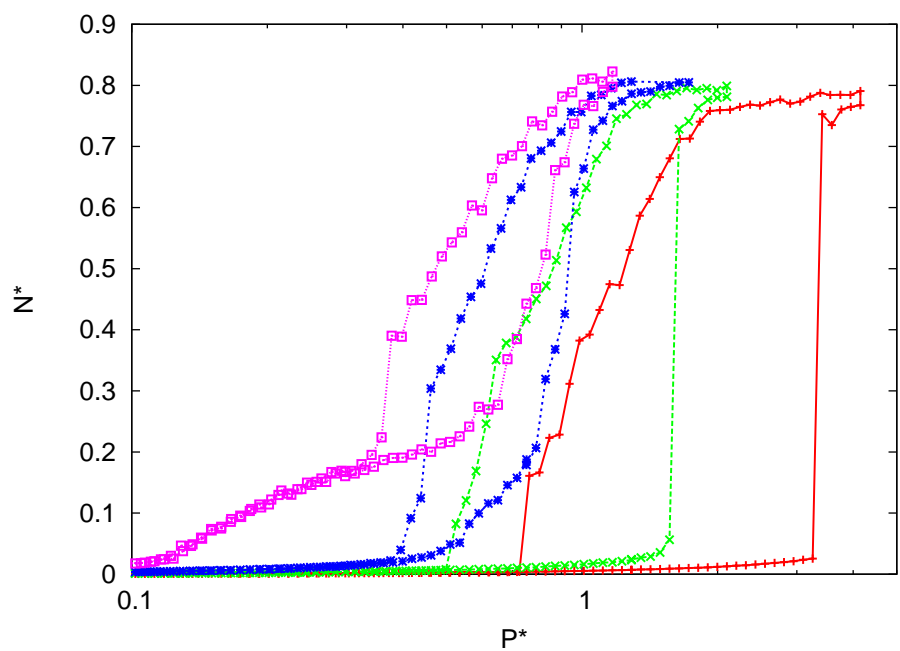

Fig. 4 (Color online) Same as Fig. 2 except R=2.5 nm

We discuss the R=1 nm case first. One observes distinct differences in Fig. 2 between the isotherms for small values of D and those for D≥ 40 meV. For large D, there occurs a continuous growth of the Ar film, outward from the wall, the analog of complete wetting behavior on planar surfaces. The dramatic increase in total density occurring (for D=50 meV) between P*=0.1 and P*=0.13 is shown in Fig. 5 to correspond to a near discontinuity in local density. In Fig. 5 we display the dimensionless local density, n = dN $\sigma^3$/(2$\pi$ r dr L), of four particular configurations. The first one (a) shows a full pore situation, with P*=0.13 and N*=0.68, just above the discontinuity in the desorption isotherm. Here there are three peaks in the density, i.e. the fluid inside the pore form three cylindrical layers, separated by an Ar interatomic distance. The second plot (b) corresponds to a slightly lower pressure, P*=0.12, but with a 30% reduced density, N*=0.5, just below the "jump" in the isotherm. For lower pressures the density decreases continuously to a bilayer (c) and single layer film (d) adsorbed on the wall of the pore. The distance from this layer to the wall is 0.17 nm, a value dictated by the characteristic scale of the interaction potential.

For small D, in contrast, there occurs relatively little initial adsorption near the wall, followed by a discontinuous, pore-filling "quasi-transition". We use this latter term to describe a nearly vertical jump, which cannot be a real thermodynamic transition for this quasi-one-dimensional system [17,18]. Another difference in behavior, dependent on D, is the presence of hysteresis: little or no hysteresis is observed to occur for D≥ 40 meV, while hysteresis is significant for D ≤35 meV. For our simulations, the triple point pressure is $P_3$=0.55 atm. We see from the data that no fluid is present at saturation (in either adsorption branch) for D< 20 meV. Thus, we identify a threshold value of D in these pores $D_{thresh}$= 20 meV. This is the analog of the recently explored threshold for wetting of planar surfaces, for which we have determined a value that we call $D_{wet}$. The analysis in Ref. [11] yields an estimate $D_{wet}$ =53 meV for Ar wetting at the triple point in the case assumed here, C=600 meV-Å$^3$. Hence, we find a ~ 60% lower threshold potential for CC in this small pore than for wetting of a planar surface. The case of a Li substrate (D=21 meV) is a borderline case for which studies of CC would be particularly interesting.

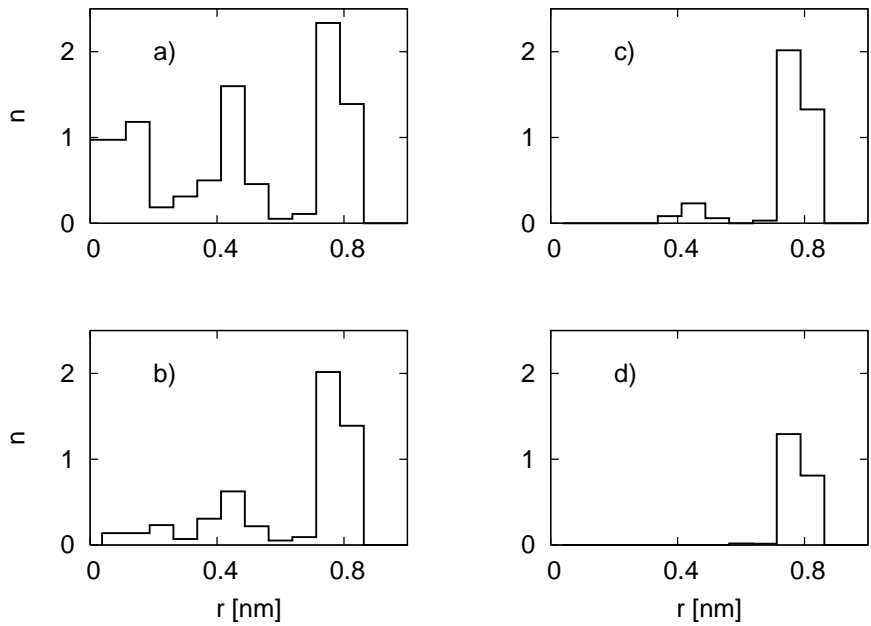

Fig. 5 Dimensionless Ar densities n(r) as a function of radial distance r for the R=1 nm pore, with D=50 meV, at various reduced densities/pressures: a) $N^*$=0.68, $P^*$= 0.13, b) $N^*$=0.5, $P^*$=0.12, c) $N^*$=0.42, $P^*$= 0.1 and d) $N^*$=0.25, $P^*$=0.04.

We turn now to the case R=1.7 nm. The results differ in some respects from those for the smaller pore. For example, the film growth near the wall is more gradual at low P for the larger pore because the attractive potential is weaker, due to the smaller curvature. While the larger pore (nearly three times as large a volume) admits more particles than the smaller pore, the relative increase in full pore capacity (larger by a factor ~2.5) is not so large as one might expect from the volume increase, a result of steric effects, i.e. the formation of shells at spacing dictated by the atomic size; see Figs. 5 and 6. One key difference in the larger pore data is the presence of hysteresis for all values of D studied here. However, the magnitude of the hysteretic jump (in N) is small in the large D case because of the significant pre-jump coating of the pore walls. Note that the CC threshold value $D_{thresh}$ =18 meV is slightly lower than that found for R=1nm. In Fig 6 we show the density decrease in the desorption isotherm, for D=50 meV. We first see CC (a) with five layers, then the density decreases continuously until $P^*$= 0.34, where it jumps discontinuously to a bilayer configuration (c). At lower pressure it evolves, as in the case of the smaller pore, to a monolayer film near the wall (d).

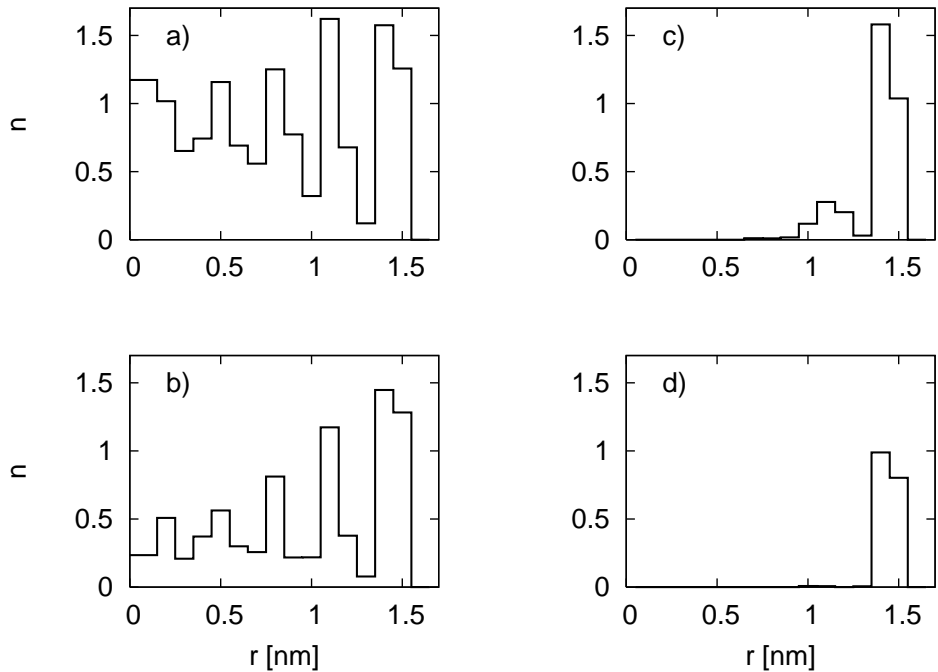

Fig. 6 Densities as a function of radial distance at various pressure for R=1.7 nm pore, with D= 50 meV: a) N*=0.77, P*=0.57, b) N*= 0.43, P*=0.34, c) N*= 0.24, P*=0.32 and d) N*=0.14, P*=0.13.

Finally, we address the largest pores considered here, R=2.5 nm. The hysteresis loops are even wider than in the previous two cases and the desorption branch shows a more gradual decrease with decreasing P. Here, we find a threshold value $D_{thresh}$ ~ 14 meV, even lower than the threshold found for the smaller pores. This finding is based on the assumption that the desorption branch is that corresponding to equilibrium. One cannot assess this assumption with our computational methods. For D=50 meV, the desorption isotherm also exhibits a discontinuity as in the case of the smaller pores (less pronounced). In Fig. 7 we can see the densities of four configurations for D=50meV. The first corresponds to a dense full pore (a), with density decreasing continuously to b) and c) and then a discontinuous decrease to a bilayer phase. Note that for the smallest pore the "jump" in N* corresponds to a quasitransition between a low density CC phase and a high density CC phase, while for the larger pores the quasitransition is between a bilayer film and a CC phase.

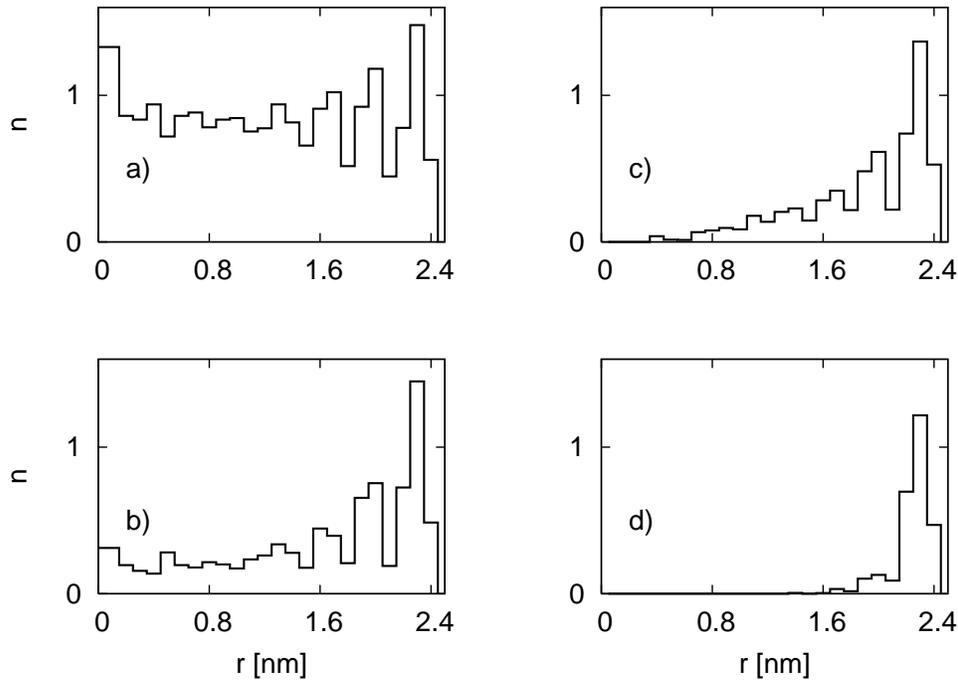

Fig. 7 Densities as a function of radial distance at various pressure for R=2.5 nm pore, with D= 50 meV: a) $N^*=0.8$, $P^*=1.0$, b) $N^*=0.44$, $P^*=0.48$, c) $N^*=0.37$, $P^*=0.43$ and d) $N^*=0.2$, $P^*=0.39$.

4. Discussion

The key results of this study pertain to the D and R dependences of the isotherms. A distinguishing characteristic of the R=1 nm data is the absence of hysteresis in the case of strongly attractive potentials, D≥ 40 meV. In contrast, hysteresis is present in our data for smaller D cases, with this radius, and for any D evaluated here for the cases R=1.7 nm and 2.5 nm. The presence of hysteresis is connected (but not inextricably linked) to the presence of CC, i.e. a jump in the coverage at a threshold value of the pressure. This connection is well known and is attributable to the presence of a free energy barrier between partially and completely full states. Note that hysteresis can also occur in cases where the pore fills continuously; see the large D data for R=2.5 nm in Fig. 4. We note that the behavior found in our simulations is that the threshold $D_{thresh}$ decreases slowly with increasing pore radius. We suspect that this trend will be reversed for larger pores, where the coordination number with the carbon atoms approaches that of the flat surface.

Our principal finding is that the vaues of the CC threshold, $D_{thresh}$ , are less than half of $D_{wet}$, the threshold values for planar surface wetting. This implies that CC can occur in a pore while the flat surface made of the same material would not be wet by the adsorbate. This finding can be rationalized for the case of a very large pore, using the so-called "simple model" of wetting. That model has been found to work well in predicting both experimental wetting behavior and simulated wetting behavior for systems undergoing wetting transitions [11]. In this model, the transition occurs when the free energy cost (involving the liquid-vapor surface tension $\sigma_{lv}$) of adsorbing a film equals the reduction in energy associated with the film-substrate interaction. The expression describing this transition for the planar surface equates the magnitude of positive and negative terms in the free energy per unit area:

$$2\sigma_{lv} = \rho \int dz \, |V(z)| \quad \text{[flat surface]} \quad [5]$$

The integral on the right is evaluated over the region of an attractive gas-substrate force ($z > z_{min}$) and $\rho$ is the film density. Here, the factor two on the left side arises because of the presence of two interfaces (film-substrate and film-vapor) involving the film. Let us compare this result with that obtained for the cylindrical pore geometry, focusing on the large R case, for which an analytic result is easily written. In that case, the surface tension cost per unit length is given by $2\pi R\sigma_{lv}$ and the magnitude of the interaction energy per unit length coincides with the expression on the right side of Eq. 5, multiplied by a circumferential factor, $2\pi R$. Hence, the CC transition criterion is

$$\sigma_{lv} = \rho \int dz \, |V(z)| \quad \text{[large pore]}$$

The factor of two difference, between this equation and Eq. 5, signifies that the well-depth required for capillary condensation is one-half that required for wetting of a flat surface. Such a result is roughly compatible with our results from numerical studies of pores having R between 1 and 2.5 nm.

In summary, we have found that $D_{thresh}$, the well-depth threshold for pore filling at saturation, is roughly one-half of $D_{wet}$, the threshold for adsorbing a thick film on a planar surface. This behavior is qualitatively consistent with the "simple model", which itself was previously found to be consistent with simulation studies on flat surfaces near the triple temperature [11]. The key difference between the two cases is attributable to the fact that the flat surface transition involves the formation of two interfaces (solid-liquid and liquid-vapor), while the CC transition involves only one.

Acknowledgment

This research has been supported by a grant from the NIRT program of NSF. We are indebted especially to Kyle Alvine, Mary J. Bojan, Oleg Gang, Lev Gelb, E. Susana Hernández, Hye-Young Kim, Peter Pershan and Nathan Urban for helpful comments.